# Analysis of the Indian ASAT test on 27 March 2019.

*Vladimir Akhmetov[1], Vadym Savanevych[2], Evgen Dikov[2]*

[1]*Laboratory of Astrometry, Institute of Astronomy, V. N. Karazin Kharkiv National University, Kharkiv, Ukraine,*
[2]*Main Astronomical Observatory NASU, Kiev, Ukraine*

**ABSTRACT**

On March 27, 2019, India tested its first anti-satellite (ASAT) missile against its own "live" satellite, Microsat-R, launched on 24 January, 2019, as part of 'Mission Shakti'. The test parameters were chosen with extreme caution to minimize the hazard of post-impact debris predicted to re-enter the earth's atmosphere within 45 days. Of the 400 pieces of debris identified by NASA, more than 60 were large enough to be tracked by the US Air Force's Space Surveillance Network and US Strategic Command's Combined Space Operations Center. Much of the information provided in the press about the ASAT missile and Microsat-R was inaccurate or misleading and did not appear to be based on scientific analysis of the data available to the public. To better understand the circumstances of this event, this paper will calculated and present detailed analysis of the tracked post-impact orbit debris from Microsat-R. All results in the current paper are based on orbital data, in the form of the NORAD two-line element (TLE) dataset from space-track.org and the own calculations of the CoLiTec group.

1. **INTRODUCTION**

Microsat-R, along with KalamsatV2, was launched from the first launch pad of Satish Dhawan Space Centre at 23:37 on 24 January 2019. The launch marks the 46th flight of PSLV. 13 minutes and 26 seconds after lift-off, the 740 kg satellite was successfully injected into the intended orbit of 274.12 km (Sun Synchronous Orbit) at an inclination of 96.567°±0.2° and a speed of 7,74 km/s. After injection, two solar arrays were deployed automatically and the ISRO Telemetry Tracking & Command Network (ISTRAC) at Bengaluru assumed control of the satellite [1].

A three-stage rocket with a diameter of 1.4 m and a length of 13.2 m was used to destroy the Microsat-R satellite. The first two stages weigh 17.2 tons in total and hold 16.7 tons of fuel. The third stage weighs 1.8 tons. The total weight of the rocket is 18.5 tons. The rocket can hit targets at altitudes up to 1000 km, as DRDO chairman Satish Reddy hinted at after the test. [2]

India notified the United States of its intent to carry out an experimental weapon test in early February, but without confirming that it would be an anti-satellite test. According to U.S. government sources with knowledge of military intelligence assessments, the United States observed a failed Indian anti-satellite intercept test attempt in February. The solid-fueled interceptor missile used during that test "failed after about 30 seconds of flight" [3]. It is unclear whether the failed February 12 test used the same missile and interceptor as the successful March 27 test.

After India's successful ASAT test on 27 March, an intense policy debate began regarding the outcome: "We have mastered anti-satellite capability and we have today shown that we can hit satellites at long ranges within a few centimeters of accuracy", stated DRDO chairman, G. Satheesh Reddy.

During a meeting with NASA employees on April 1, NASA Administrator, Jim Bridenstine, delivered sharply critical remarks about India's March 27 anti-satellite weapons (ASAT) test [4]. Calculations by NASA and DOD after the ASAT Indian test found that the risk of debris striking the ISS went up by 44 percent over a ten-day period following the test, Bridenstine told NASA employees. "That is a terrible, terrible thing, to create an event that sends debris in an apogee that goes above the International Space Station," Bridenstine declared. "And that kind of activity is not compatible with the future of human spaceflight that we need to see happen. We are charged with commercializing low Earth orbit. We are charged with enabling more activities in space than we've ever seen before, for the purpose of benefiting the human condition… all of those are placed at risk when these kinds of events happen. And if one country does this, then other countries feel they have to do it, too. It's unacceptable, and NASA needs to be very clear about what its impact to us is." [5].

From India's perspective, the ASAT test is also an important achievement that boosts national prestige. Prime Minister Narendra Modi promoted the ASAT test as a sign of India's entry into a select club: 'The launch under Mission Shakti has put our country in the space super league.'

With a national election coming up, and Modi promoting his defense and national security credentials, especially after the Indian response to Pakistan following the Pulwama attack, the ASAT test plays to his domestic political agenda, as much as it seeks to emphasize India's prestige globally. [6]

## 2. MICROSAT-R SATELLITE FRAGMENT ANALYSIS

At the end of February and early March, Microsat-R carried out several orbit corrections. By March 27, 2019, Microsat-R had reached an almost circular orbit with a perigee of 260 *km* and apogee of 282 *km*. [7]

Using proprietary software developed by the CoLiTec group the orbital elements of all fragments were calculated based on two-line element (TLE) data of NORAD using space-track.org online resource.

We can calculate the values of impulse and corresponding energies using the orbital parameters of Microsat-R fragments that formed by "hit satellite with a few centimeters of accuracy". The longitude changes of the ascending node (*ΔΩ*) and inclination (*Δi*) in one turn are determined using the following equations:

$$\begin{cases} \Delta\Omega = \dfrac{1}{V_{sat}} \dfrac{\sin(U)}{\sin(i)} \Delta V_w; \\ \Delta i = \dfrac{1}{V_{sat}} \cos(U) \Delta V_w, \end{cases}$$

where *Vsat* is the circular velocity of the satellite; *U* is the argument of latitude (angular distance of the satellite position to the ascending node), *ΔVw* is the component impulse perpendicular to the orbit plane. In our case, the orbital inclination of the Microsat-R satellite is *i = 96.567 degrees*, therefore the *sin(i)* is equal to approximately 1, also, the largest possible value of the functions *sin(U)* and *cos(U)* is equal to 1. For the change in longitude of the ascending node (*ΔΩ*) and inclination (*Δi*), impulse have the order *ΔVw/Vsat*. To rotate the orbit plane by one radian (57°.3), significant impulse *ΔVw* equal to the velocity of the satellite is required. Rotation of the orbit plane (change of longitude of the ascending node and inclination) by just one angular degree will require the application of an impulse $\Delta Vw = 140$ m/s, which indicates the extreme difficulty of turning the plane of orbit. Changing the orbit plane is one of the most energy expensive maneuvers in terms of fuel.

Considering the change parameters of satellite motion which determine the form and size of trajectory in the orbit plane, we researched the parameter (*p*) that characterizes the linear dimensions of the orbit and the eccentricity characterizing the shape of the orbit:

$$\begin{cases} \Delta p = \dfrac{T}{\pi} \Delta V_t; \\ \Delta e = \dfrac{1}{V_{sat}}(\Delta V_s \sin(v) + \Delta V_t 2\cos(v)), \end{cases}$$

where *T* is the orbital period of the satellite around the Earth; *v* is the true anomaly (defines the position of the orbit satellite along the ellipse). Changing the shape of the orbit is a very economical maneuver. Thus, an impulse of *140 m/s*, which would be enough to rotate the orbit just 1 degree, is sufficient to change the altitude by *240 km*. [8]

Using tracking TLE data and orbit calculations for the Microsat-R fragments made by CoLiTec group, we calculated the *ΔV* (impulse to the original Microsat-R velocity) necessary to achieve corresponding orbits, taking into account changing orbital altitudes, longitude of the ascending node, and inclination. Each point on the figures corresponds to one fragment of the Microsat-R satellite tracked via the TLE dataset on space-track.org.

The Figure shows the distribution of the value apogee and perigee (*km*) of the Microsat-R fragments. The blue dot shows the corresponding apogee and perigee of Microsat-R until March 27. As can be seen from the figure, most of the fragments moved to higher orbits with an apogee of 300 to 2265 *km*. This indicates a significant impulse in the direction of satellite motion.

All Microsat-R fragments with an impulse inverse the motion of the satellite moved to a lower orbit and "burned" in the Earth's atmosphere within a few days. According to the results of NORAD observations, we can only carry out calculation for the Microsat-R satellite fragments that received the impulse along and perpendicular plane and are still in orbit. Unfortunately, we cannot estimate the individual mass of the Microsat satellite fragments, therefore we calculate the impulses and energies per one kg. Figure 2 shows the value of impulses (m/s/1kg) which obtained the Microsat-R fragments in the direction of satellite motion (*X* axis) and also perpendicular to orbit plane (*Y* axis). These value of impulses were calculated from an increase of the value of orbital apogee and perigee and a rotation the orbit plane respectively.

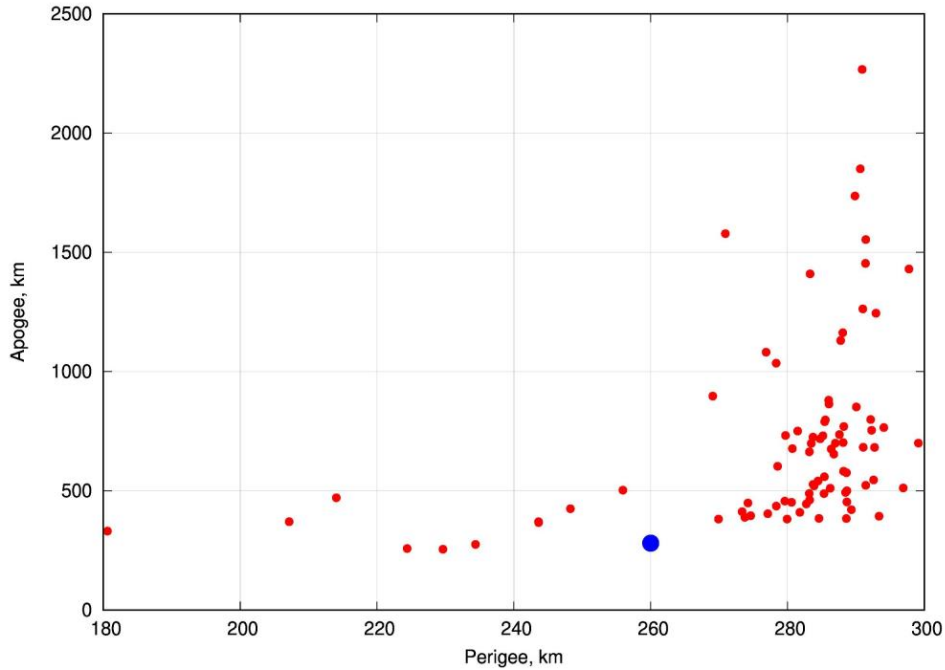

*Figure 1: Distribution of the value apogee and perigee (km) of the Microsat-R fragments.*

The resulting distribution of impulse *ΔV* values range from -50 to 1200 *m/s/1kg* into direction of satellite motion and from -1000 to 2000 *m/s/1kg* in the direction perpendicular to the orbit plane.

Figure 3 shows the total amount of energy obtained by the Microsat-R satellite fragments, depending on the value of the apogee. The trend line shows the mean value of energy for a possible scatter of Microsat-R fragments as a function of apogee. The mean value of energy obtained for all fragments is almost unchanged from their orbit altitude. The majority of energy is spent on rotation along the orbital plane by Microsat-R fragments that make small changes in orbit altitude. We can say that Microsat-R satellite fragments obtained the same impulse in all directions, which may indicate the undermining of the satellite without a head-on collision with a rocket.

As documented in the literature, in the case of cubic charge explosion in a vacuum, six mutually perpendicular rays are formed instead of a spherical front [9]. In the published IR video we can see the instant of rocket collision with the Microsat-R satellite [10]. Figure 4 clearly shows 4 rays, 2 which remain directed perpendicular to the observation plane. This confirms the above version of the explosion. However, according to the available data, it is impossible to precisely determine the origin of the explosion. Explosion could be the result of fuel detonation due to the collision of the ASAT rocket with a satellite, or an internal remote explosion unrelated to the ASAT rocket impact.

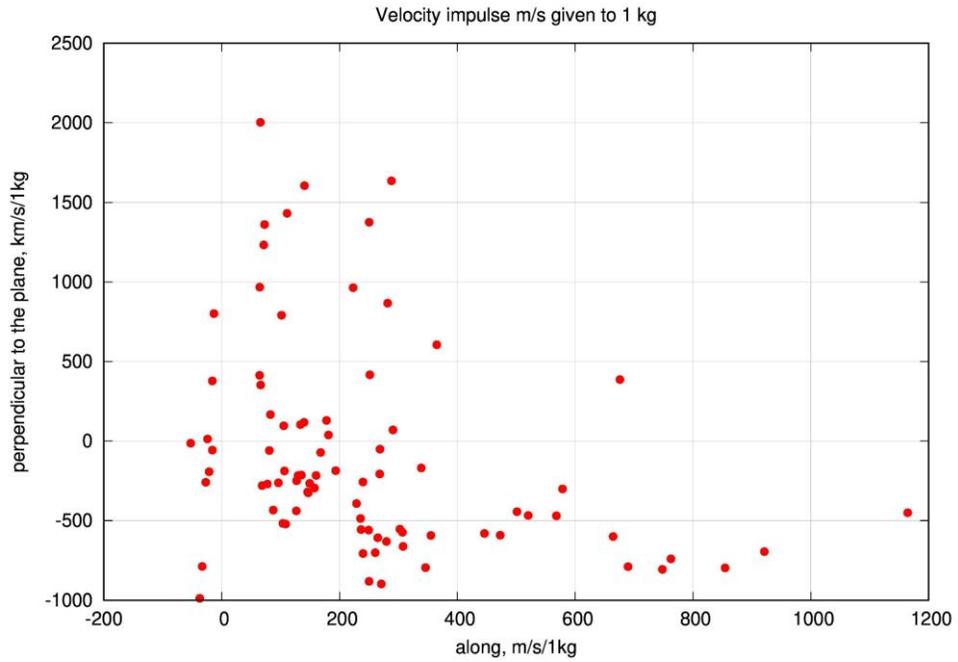

*Figure 2: Distribution of the value of impulses obtained by the Microsat-R fragments in the direction of satellite motion (X axis) and perpendicular to the orbit plane (Y axis).*

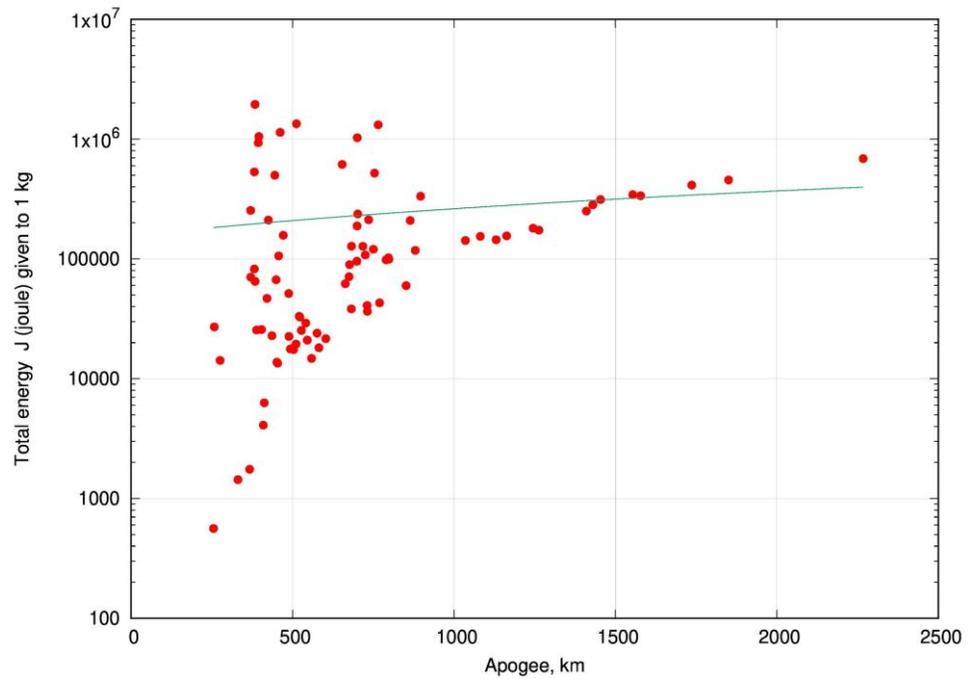

*Figure 3: Total energy obtained by Microsat-R fragments, depending on the apogee value.*

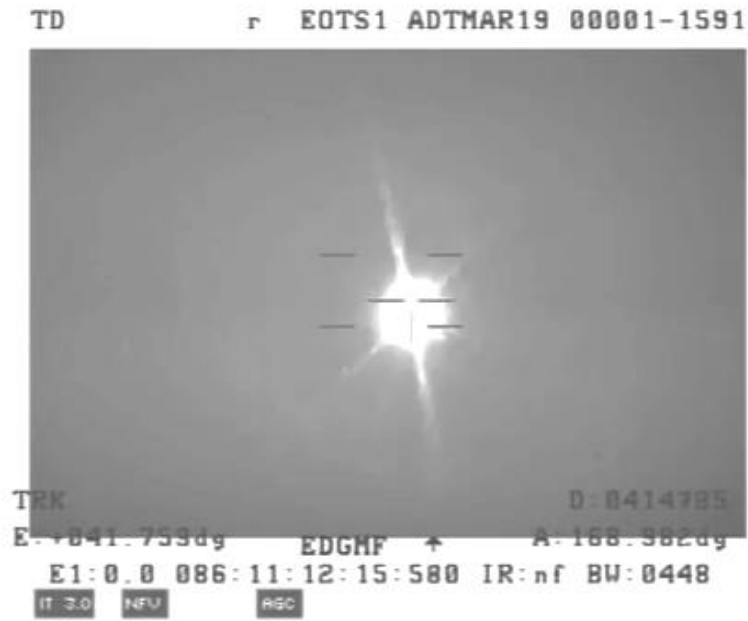

*Figure 4: Screenshot from the video of the ground-based IR camera taken during a rocket collision with the Microsat-R satellite*

### 3. CONCLUSION

Using proprietary software developed by the CoLiTec group, the orbit elements for all fragments tracked by NORAD were calculated via the TLE dataset on space-track.org. The results show that after the "head-on collision" of the Microsat-R satellite with the anti-satellite rocket at an altitude of approximately 270 km had been declared, NORAD published information on 83 newly registered fragments of space debris on 10 May 2019.The apogee of 69 of these fragments was above the ISS orbit, at altitudes of 400 to 2265 km. According to orbit parameters of the Microsat-R satellite fragments, impulses and corresponding energies were calculated. Analysis of IR video of the rocket collision with the Microsat-R satellite was carried out. All these results confirm the version of explosion of Microsat-R satellite. However, with the available data, it is impossible to confirm the origin of the explosion.

The results of our research showed the significant differences from expected "head-on collision" of the rocket with Microsat-R satellite, as a result of which all the fragments had been to burn in the atmosphere within 45 days. We see the unacceptable for human activity in space - the 'hard kill' of the satellite, the resulting explosion and the significant change of the satellite fragments orbit parameters. It would therefore be wise to sideline the 'hard kill' approach and develop techniques for a 'soft kill' that avoids the risk of 'Kessler syndrome' - event.

### REFERENCE

1. https://www.reddit.com/r/ISRO/comments/aglwrb/pslvc44_microsatr_mission_updates_and_discussion/
2. http://www.indiandefensenews.in/2019/04/all-you-need-to-know-about-pdv-mk-ii.html

3. https://thediplomat.com/2019/04/exclusive-india-conducted-a-failed-anti-satellite-test-in-february-2019/
4. https://arstechnica.com/science/2019/03/india-shoots-down-a-weather-satellite-declares-itself-a-space-power/
5. https://arstechnica.com/tech-policy/2019/04/india-asat-test-debris-poses-danger-to-international-space-station-nasa-says/
6. https://www.aspistrategist.org.au/will-indias-anti-satellite-weapon-test-spark-an-arms-race-in-space/
7. https://heavens-above.com/OrbitHeight.aspx?satid=43947&startMJD=58484.0
8. http://12apr.su/books/item/f00/s00/z0000023/st051.shtml
9. https://sites.wrk.ru/sites/net/zh/zhurnalko/images/1/b/1baaf6583d7dad512f75/page0006.jpg
10. https://www.youtube.com/watch?v=1-J4jVlx5Do&feature=youtu.be